# Large Proximity-Induced Spin Lifetime Anisotropy in Transition Metal Dichalcogenide/Graphene Heterostructures

Talieh S. Ghiasi[1*], Josep Ingla-Aynés[1], Alexey A. Kaverzin[1], Bart J. van Wees[1]

**Van-der-Waals heterostructures have become a paradigm for designing new materials and devices, in which specific functionalities can be tailored by combining the properties of the individual 2D layers. A single layer of transition metal dichalcogenide (TMD) is an excellent complement to graphene (Gr), since the high quality of charge and spin transport in Gr is enriched with the large spin-orbit coupling of the TMD via proximity effect. The controllable spin-valley coupling makes these heterostructures particularly attractive for spintronic and opto-valleytronic applications. In this work, we study spin precession in a monolayer MoSe$_2$/Gr heterostructure and observe an unconventional, dramatic modulation of the spin signal, showing one order of magnitude longer lifetime of out-of-plane spins compared with that of in-plane spins ($\tau_\perp \approx 40$ ps and $\tau_\parallel \approx 3.5$ ps). This demonstration of a large spin lifetime anisotropy in TMD/Gr heterostructures, is a direct evidence of induced spin-valley coupling in Gr and provides an accessible route for manipulation of spin dynamics in Gr, interfaced with TMDs.**

Graphene with its high charge carrier mobility and weak spin-orbit coupling (SOC) is an excellent host for long-distance spin transport[1-4]. However, data storage and information processing in spin-based devices require active control of the spin degree of freedom[5]. Therefore manipulation of the spin currents, i.e. tuning spin-polarization and spin lifetime, has been a topic of recent theoretical[6-9] and experimental[10-15] research. To fulfill this goal, one of the main approaches is fabrication of hybrid devices, in which the properties of the 2D building blocks complement each other[16,17]. The strong SOC of TMDs, orders of magnitude larger than the one of Gr[18], can modulate the spin dynamics in the Gr channel, while the high quality of charge transport of Gr is preserved[15]. The induced SOC in Gr via proximity effect of TMD can be in the order of 10 meV[7], experimentally confirmed by the observation of weak anti-localization[13,14] and spin Hall-effect[15] in these heterostructures, and by the suppression of the spin lifetimes[11,12] to less than few picoseconds.

The modulation of spin currents in bulk TMD/Gr vdW heterostructures has been already reported[10,11] based on gate-controlled spin absorption by the TMD. Moreover, the spin-valley locking has been used for optical excitation of spins[19,20] in TMD, which are injected into and transported by the underlying Gr. However, in this work we study how the spin transport properties of Gr are influenced by the proximity of monolayer TMD.

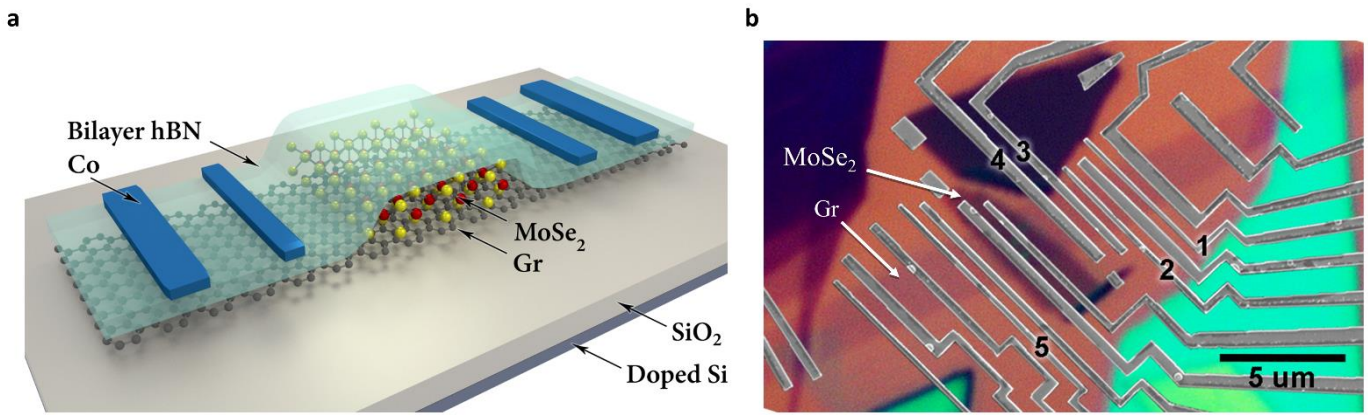

**Figure 1| Device geometry.** (a) Sketch of the MoSe$_2$/Gr vdW heterostructure on a SiO$_2$/Si substrate with a top layer of bilayer hBN, used as a tunnel barrier for spin-injection and detection in Gr with Co contacts. (b) Combination of an optical microscope (OM) image of the vdW heterostructure and an SEM image of the Co contacts. The green flake is bulk hBN. The used electrodes for measurements are numbered. The width of the Gr channel is about 2.4 μm.

[1] Zernike Institute for Advanced Materials, Physics of Nanodevices, University of Groningen, 9747 AG Groningen, The Netherlands
*email: t.s.ghiasi@rug.nl

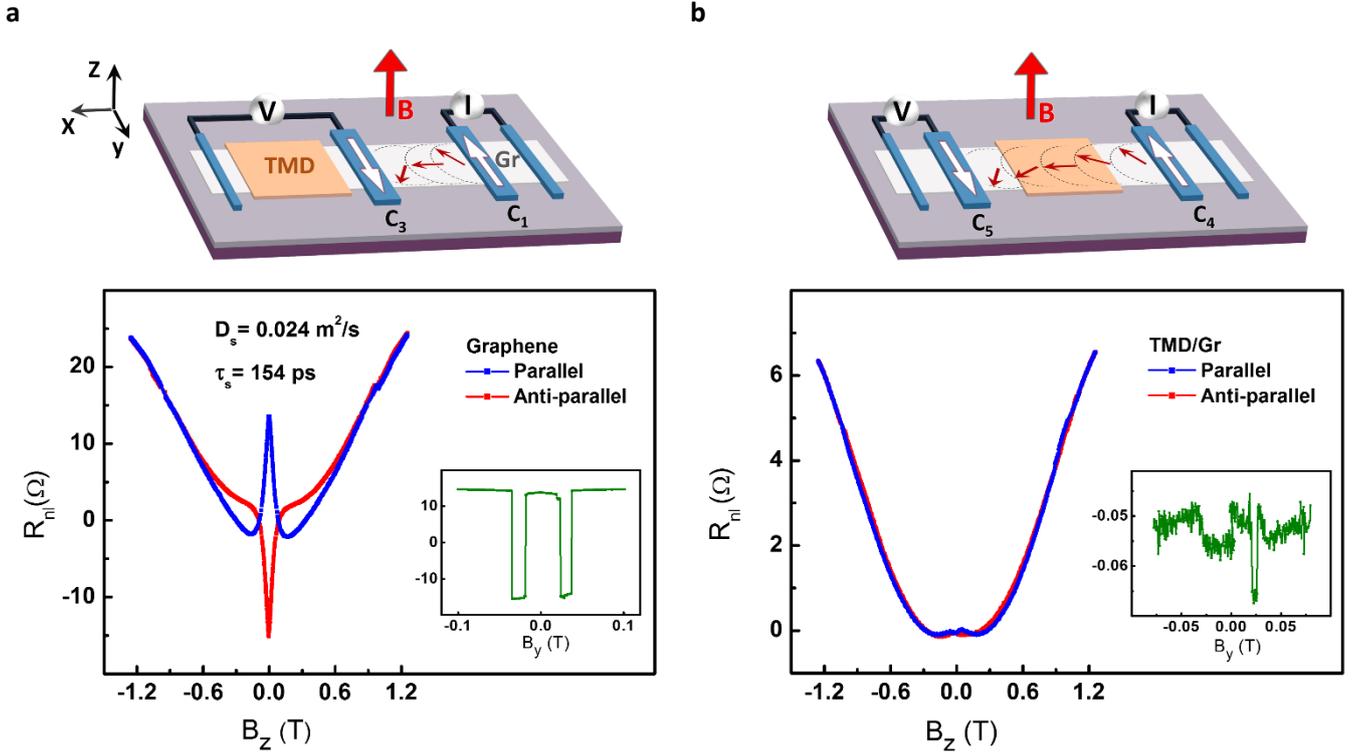

**Figure 2| Comparison of Hanle precession measurements with out-of-plane magnetic field ($B_z$).** The device sketches with nonlocal measurement geometries are illustrated (contacts are numbered according to Figure 1b). Nonlocal magnetoresistance ($R_{nl}$) as a function of $B_z$ and the corresponding nonlocal spin-valve (shown in the inset) are measured (a) on the Gr channel (length= 2.1 μm) with $C_1$ as the spin injector and $C_3$ as the detector and (b) across the TMD/Gr region with $C_4$ as the spin injector and $C_5$ as the detector (channel length is 4.1 μm, covered with 2 μm $MoSe_2$). In our measurement setup, $B_z$ is limited to 1.2 T, which is not sufficient for complete out-of-plane saturation of the contact magnetization. Therefore, the reported magnitude for the out-of-plane spin signal is the lower bound of the real value.

We address the induced anisotropic nature of spin relaxation in these hybrids, originating from the strongly coupled spin and valley degrees of freedom[21]. Our results, consistent with the theoretical predictions[8], give an insight into the valley-coupled spin dynamics in TMD/Gr heterostructures and are very relevant to acquire a complete understanding of the physics of (opto-) valleytronics and spintronics in these systems.

We fabricate devices based on a vdW heterostructure of monolayer $MoSe_2$/monolayer Gr, covered with an hBN bilayer (Figure 1). These atomically thin layers are exfoliated from their bulk crystals and are stacked by a dry pick-up technique[22] that provides high quality and polymer-free interfaces. To study spin-transport, we use ferromagnetic cobalt contacts using the bilayer hBN as a tunnel barrier that allows for highly efficient electrical spin injection and detection[23].

The conventional four-terminal non-local geometry[24] for injection and detection of pure spin currents is shown in Figure 2a. We measure the non-local resistance ($R_{nl}$= V/I) while sweeping the magnetic field ($B_y$) along the easy axis of the Co contacts. All the measurements are carried out at 75 K. The spin-valve signal is defined as the difference in $R_{nl}$, measured in parallel and antiparallel magnetization configurations of the contacts ($\Delta R_{nl} = R_p - R_{ap}$). The $\Delta R_{nl}$ signal of 30 Ω is measured over 2.1 μm of the Gr channel. When an out-of-plane magnetic field ($B_z$) is applied, the spins undergo Larmor precession in the x-y plane while diffusing. By measuring $R_{nl}$ as a function of $B_z$, we acquire the so-called Hanle precession curves. The spin lifetime ($\tau_s$), diffusion coefficient ($D_s$) and contact polarization (~40%) are obtained by fitting the Hanle curves to the solution of Bloch equations[25]. Interestingly, increasing $B_z$ beyond the typical fields sufficient for significant spin dephasing, enhances $R_{nl}$ over its value at B = 0 T. This increase in the spin-signal is attributed to the contribution of the out-of-plane spins when the magnetization of the Co electrodes gets pulled out of the graphene plane. We observe that the ratio of the spin signal at high $B_z$ over the in-plane spin signal (at $B_z$ = 0 T) increases as the detector contact approaches the TMD/Gr region (Supplementary S4.1). This observation indicates that the neighboring TMD/Gr region acts as a spin sink for the in-plane spins, thereby suppressing their measured relaxation time compared with that of out-of-plane spins. In order to further understand the effect of the monolayer $MoSe_2$ on spin-transport in Gr, we measure $R_{nl}$ across the TMD/Gr region. In Figure 2b, the spin-valve measurement shows a considerable suppression of the in-plane spin signal to ≈ 15 m$\Omega$, which is about 300 times smaller than the in-plane spin signal in pristine Gr with the same channel length. When we apply $B_z$ and the z-component of the magnetization of the Co contacts increases, we observe that $R_{nl}$ enhances dramatically to values over 6 $\Omega$. This observation implies that the out-of-plane spin signal exceeds the in-plane signal by about three orders of magnitude. Such a giant contrast between the in- and out-of-plane spin signal can only be understood if the in-plane spin lifetime in the TMD/Gr heterostructure is shorter than 3.5 ps, two orders of magnitude smaller than that of the pristine Gr channel (Supplementary S4.2).

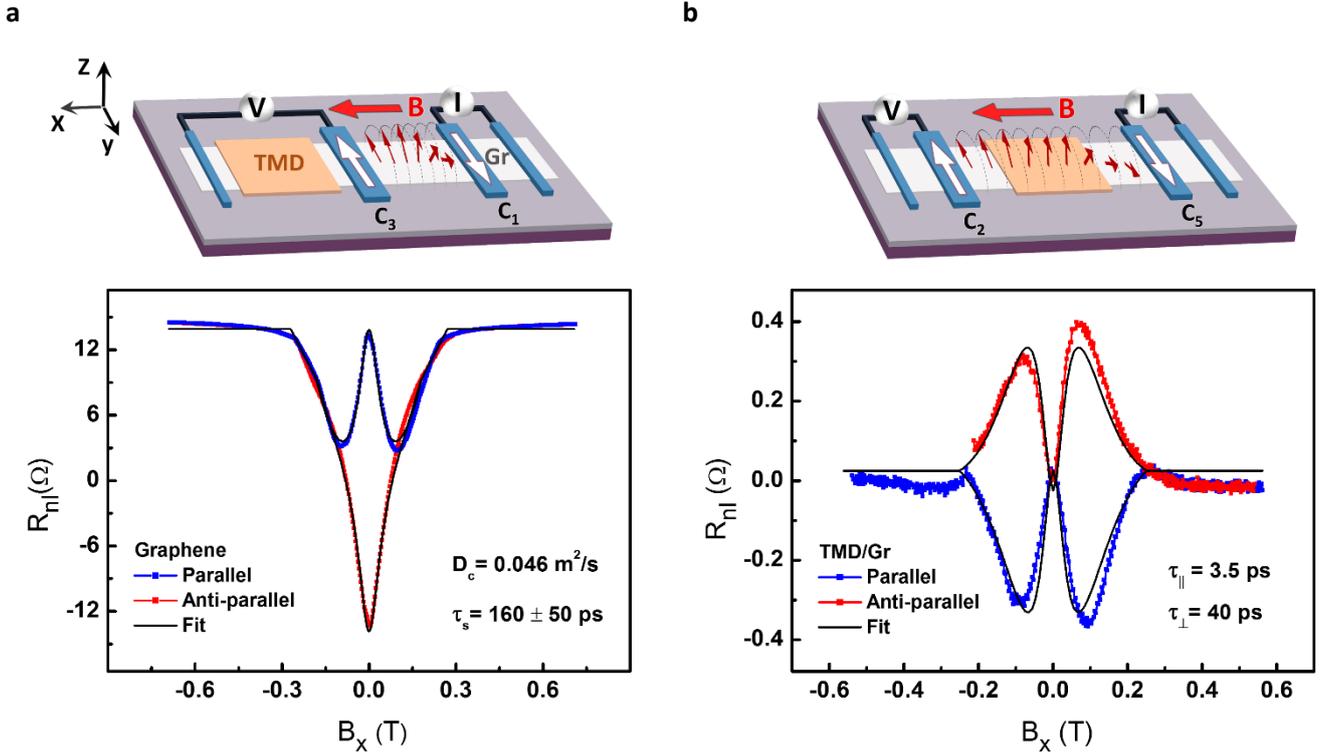

**Figure 3 | Comparison of Hanle precession measurements with in-plane magnetic field ($B_x$).** The device sketches with nonlocal measurement geometries are illustrated (contacts are numbered according to Figure 1b). $R_{nl}$ as a function of $B_x$ is measured (a) on the Gr channel, fitted with the uniform model (spin injector is $C_1$ and detector is $C_3$) and (b) across the TMD/Gr region, with a fit by the four-region model (spin injector is $C_5$ and detector is $C_2$) (channel length is 5.6 $\mu$m, covered with 2 $\mu$m MoSe$_2$). The fit to the data is obtained for $\tau_\parallel$ = 3.5 ps and $\tau_\perp$ = 40 ps. Note that in (b), the sketch of the sample is rotated by 180 degrees. See supplementary for details of uniform and four-region models.

To get more information regarding the lifetimes of in-plane spins ($\tau_\parallel$) and out-of-plane spins ($\tau_\perp$), we apply an in-plane magnetic field ($B_x$). This direction of magnetic field makes the spins precess in the y-z plane, thus probing both $\tau_\parallel$ and $\tau_\perp$[26]. By measuring $R_{nl}$ in the pristine Gr regions, we obtain the Hanle precession curves in Figure 3a. As expected, the in-plane spin-signal has its highest value at B= 0 T and decreases when the spins precess. Sufficiently large $B_x$ aligns the Co electrode magnetization in the x direction (saturating at $B_x \approx$ 0.3 T) and therefore the spin signal is restored to its initial value (at B= 0 T). The behavior of the contacts is explained by using the Wohlfarth-Stoner model[27]. The spin transport parameters can be extracted by the fit to the solutions of the Bloch equations[25] which confirms the isotropic spin relaxation in the Gr region (Supplementary S4.3).

However, when we measure $B_x$ induced Hanle precession across the TMD/Gr region, we observe a dramatically different behavior (Figure 3b). At B= 0 T, the value of $R_{nl}$ is small and is caused by the in-plane spin transport (in the y-direction). As $B_x$ increases the in-plane spins start to precess in the y-z plane which generates out-of-plane spins that have longer lifetime. Therefore, the $R_{nl}$ considerably increases in magnitude to about 35 times larger values (at $B_x \sim$ 0.1 T) and reverses sign. Beyond 0.1 T the signal decreases due to the both spin dephasing and saturation of contact magnetization, recovering the in-plane spin signal. This observation is again a direct proof of anisotropic spin-transport in TMD/Gr heterostructure. Using the Bloch equations, we can fit the data to an anisotropic relaxation model[28], accounting for the difference between $\tau_\parallel$ and $\tau_\perp$ in the TMD/Gr region (see

Supplementary S4.4). A fit to the data is obtained with values of $\tau_\parallel = 3.5$ ps and $\tau_\perp = 40$ ps, confirming the large spin lifetime anisotropy in the Gr induced by the monolayer MoSe$_2$. The measurements (shown in Figure 3) are carried out at zero gate voltage (Vg), with carrier densities of 4.5 x 10$^{12}$ cm$^{-2}$ and 1.8 x 10$^{12}$ cm$^{-2}$ in the pristine Gr and TMD/Gr regions, respectively. With the change of Vg (to ±40 V) we do not see any considerable change in the anisotropy, indicating that the spin absorption cannot be the dominant mechanism for spin-relaxation in TMD/Gr region (Supplementary S5).

According to the recent theoretical predictions[8], the dynamics of spin transport in Gr in proximity of TMD are governed by the Dyakonov-Perel (DP) mechanism, which plays a major role in systems with broken inversion symmetry. In a TMD/Gr heterostructure, the strong spin-valley coupling of the TMD is imprinted onto the Gr channel and controls the dynamics of the in-plane spins. The in-plane spin relaxation is affected by both intervalley and momentum scattering, but the former is dominant ($\tau_\parallel \propto 1/\tau_{iv}$, with $\tau_{iv}$ the intervalley scattering time). In contrast, the out-of-plane spin relaxation is mainly controlled by the momentum scattering in this system ($\tau_\perp \propto 1/\tau_p$, with $\tau_p$ the momentum relaxation time). The spin lifetime anisotropy can be calculated by:

$$\frac{\tau_\perp}{\tau_\parallel} = \left(\frac{\lambda_{VZ}}{\lambda_R}\right)^2 \frac{\tau_{iv}}{\tau_p} + \frac{1}{2}$$

where $\lambda_{VZ}$ and $\lambda_R$ are the valley Zeeman and Rashba spin-orbit coupling constants[8]. These terms have been calculated by first principles as $\lambda_{VZ}$= -0.175 meV and $\lambda_R = 0.26$ meV for a MoSe$_2$/Gr heterostructure[7]. Our experimental observation of spin lifetime anisotropy estimated as $\tau_\perp/\tau_\parallel \sim 11$, corresponds to an intervalley scattering time of $\tau_{iv} \sim 23\ \tau_p \approx 2$ ps (with $\tau_p$= 0.076 ps). This value matches well with the reported range for $\tau_{iv}$ in Gr, extracted from weak localization measurements[13,29].

In conclusion, we have reported the first direct observation of the spin lifetime anisotropy in a TMD/Gr heterostructure, consistent with the theoretical predictions of the TMD SO-induced proximity effects in Gr[8]. The estimated out-of-plane spin relaxation time is one order of magnitude larger than that of the in-plane spins. This result is explained by considering the dominant role of the inter-valley scattering in the relaxation of the in-plane spins. The effect is attributed to the spin-valley coupling in TMD/Gr as a consequence of the strong spin-orbit coupling in TMD, the significant wavefunction overlap between Gr and the TMD, and the associated inversion symmetry breaking. We have demonstrated that the manipulation of spin-transport in the Gr channel is possible by controlled stacking of atomically thin building blocks, where unprecedented insights into the nature of the spin-orbit interactions are provided by a simple and novel approach in the spin precession experiments.

## Methods

**Device Fabrication.** Graphene, MoSe$_2$ and hBN flakes are mechanically exfoliated from their bulk crystals (supplier: HQ Graphene) onto a SiO$_2$/doped Si substrate. We identify the monolayers of Gr and MoSe$_2$ and the bilayer of hBN by their optical contrast with respect to the substrate[30]. To confirm the thickness of the layers, we performed AFM height profile measurements (Figure S1). Subsequently, the bilayer hBN and monolayer MoSe$_2$ flakes are transferred on top of Gr on a SiO$_2$ (300 nm)/Si substrate by a polymer-based dry transfer technique using PC (Poly(Bisphenol A)carbonate) and a PDMS stamp[22]. This technique, followed by cleaning steps of dissolving the PC in chloroform and 12 hours of annealing in Ar/H$_2$ flow at 350 °C, provides us with high quality vdW heterostructure (bilayer hBN/monolayer MoSe$_2$/monolayer Gr). Electrodes are patterned by electron-beam lithography on PMMA e-beam resist. Device fabrication is accomplished with evaporation of 65 nm Cobalt by e-beam evaporator (at a pressure of 5.0×10$^{-7}$ mbar) and Lift-off in 30 °C acetone.

**Electrical Measurements.** Charge and spin transport measurements are performed in vacuum ($\sim 10^{-7}$ mbar) at 75 K, using a standard low-frequency (< 20 Hz) lock-in technique and AC current of 100 nA to 5 µA. For charge transport measurements, we have used a local four-terminal geometry in order to eliminate the effect of the contact resistances. Detection of pure spin signal is done by using a non-local four-terminal geometry, illustrated in the main manuscript.

**Acknowledgements**

We kindly acknowledge S. Roche, J. Fabian and G. E. W. Bauer for insightful discussions. We would like to thank T. J. Schouten, H. M. de Roosz, J. G. Holstein, and H. Adema for technical support. Also we acknowledge J. Peiro and M. Gurram for their assistance. This research has received funding from the Dutch Foundation for Fundamental Research on Matter (FOM), as a part of the Netherlands Organisation for scientific Research (NWO), FLAG-ERA, the People Programme (Marie Curie Actions) of the European Union's Seventh Framework Programme FP7/2007-2013/ under REA grant agreement No. 607904-13 Spinograph, the European Union Seventh Framework Programme under grant agreement No. 604391 Graphene Flagship and supported by NanoLab NL.

**Author contribution**

T.S.G. fabricated the samples and carried out the measurements with assistance of A.A.K. and J.I-A. Modelling was done by J.I-A and A.A.K. with input from B.J.v.W. Data analysis was done by T.S.G., J.I-A. ,A.A.K. and B.J.v.W., who supervised the project. The paper is written by contribution of all authors.

**Additional information**

Correspondence and requests for materials should be addressed to T.S.G.

**Competing financial interest**

The authors declare no competing financial interest.

# Supplementary Information

## Large Proximity-Induced Spin Lifetime Anisotropy in Transition Metal Dichalcogenide/Graphene Heterostructures

Talieh S. Ghiasi[2*], Josep Ingla-Aynés[1], Alexey A. Kaverzin[1], Bart J. van Wees[1]

**Table of contents:**

[2] Zernike Institute for Advanced Materials, Physics of Nanodevices, University of Groningen, 9747 AG Groningen, The Netherlands
* email: t.s.ghiasi@rug.nl

## S1. AFM characterization of the vdW heterostructure

The height profile of the bilayer hBN/monolayer MoSe$_2$/monolayer Gr heterostructure on a SiO$_2$/Si substrate is obtained by the ScanAsyst mode of Atomic Force Microscopy (AFM, Bruker Multi Mode 8), shown in AFM image of Figure S1a. We attribute the light spots, probed on the region of Gr and MoSe$_2$ flakes, to the conventional bubbles formed at the interfaces between the 2D materials[1]. Van der Waals forces bring the adjacent layers in close interaction with each other and squeeze out the trapped molecules (adsorbates, present on the individual layers before transfer, e.g. water molecules, hydrocarbons). Thereby, accumulation of the adsorbate molecules at the interfaces forms submicron-size bubbles. Observation of these bubbles can be an indication of proper adhesion between layers of vdW heterostructure, meaning that the area between bubbles is an adsorbate-free, perfectly clean interfacial region. Recent studies[1] on bubbles formed in Gr, MoS$_2$ and hBN 2D crystal layers conclude that the shape and internal pressure of the bubbles is determined by the competition between van der Waals attractions and the elastic energy needed for deformation of the crystal layer. Therefore, the stiffness of the 2D crystal influences the size of the bubbles, leading to smaller and more scattered ones. Figure S1b compares the height profiles for each layer of the vdW heterostructure. The thickness of the monolayer Gr, bilayer hBN and monolayer TMD are measured about 0.45 nm, 0.7 nm and 0.8 nm, respectively. These values are in agreement with the reported thicknesses, measured by AFM on SiO$_2$/Si substrates[2,3,4].

## S2. Charge transport characterization

All the electrical measurements are performed in vacuum ($\sim 10^{-7}$ mbar) at 75 K, using a standard low-frequency (< 20 Hz) lock-in technique and AC current of 100 nA to 5 µA.

We perform local four-terminal measurements to evaluate the charge transport properties of Gr and TMD/Gr channels (Figure S2). These measurements show n-type doping of the Gr channel in our device. We can calculate the electron mobility (µ) in Gr, with equation (S1)[6] from the conductivity as a function of charge carrier density (n) which includes both long and short-range scatterings.

$$\sigma = \left(\frac{1}{ne\mu+\sigma_0} + \rho_s\right)^{-1} \tag{S1}$$

In this equation, $\rho_s$ is the contribution to resistivity from short-range scattering and $\sigma_0$ is the residual conductivity at the charge neutrality point. The extracted values for the electron mobility at 75 K are about $10{,}000 \text{ cm}^2/\text{V} \cdot \text{s}$ and $5{,}600 \text{ cm}^2/\text{V} \cdot \text{s}$ in the Gr and TMD/Gr regions, respectively. We can calculate the momentum scattering time ($\tau_p$) from the charge transport coefficients, using equation[7]

$$\tau_p = h\sigma/(2e^2 v_F \sqrt{(\pi n)}) \tag{S2}$$

where h, $v_F$ and n are the Plank's constant, Fermi velocity and the density of charge carriers, respectively. We obtain the values of $\tau_{p,Gr} = 0.09$ ps and $\tau_{p,TMD/Gr} = 0.076$ ps at Vg= 0 V for momentum scattering in Gr and TMD/Gr regions, respectively.

From the charge transport measurements, it is possible to obtain information on the band alignment of TMD and Gr in these heterostructures, that is one important parameter that should be taken into account to design these devices. Based on the local four-terminal charge transport, the reported band alignment of Gr/MoS$_2$[8] and the calculated band-offsets of MoS$_2$ and MoSe$_2$[9], we conclude that the Fermi level of the (n-doped) Gr channel is within the band gap of the monolayer MoSe$_2$. Theoretical studies have shown that because of the insignificant dipole formation at the TMD/Gr interface, the valid level alignment can be obtained by aligning the vacuum levels of the isolated layers[10]. Moreover, the presence of the bilayer hBN as tunnel barrier layer between MoSe$_2$ and Co electrode eliminates chemical interaction between the metal contact and TMD, thus reduces the Fermi-level pinning at the interface.

In these heterostructures, due to the absence of the dangling bonds on the exfoliated layers of Gr and MoSe$_2$, formation of sharp and high quality atomic interfaces is expected. However, each material preserves its individual electronic properties due to the weak interaction between the layers[3].

**Electrical contact resistances**

By three-terminal measurements, we evaluate the resistance of the electrical contacts with bilayer hBN tunnel barriers. The contact resistances on hBN/Gr vary in the range of 3 to 11 kΩ at RT and the range of 5 to 13 kΩ at 75 K. These values guarantee the impedance matching for the spin transport measurements and, therefore, negligible back-flow induced spin-relaxation[11]. The average resistance area product (RA) of these contacts is estimated as 6.6 and 11.4 kΩ·µm$^2$ at RT and 75 K, respectively. The range of the contact resistance, together with the high efficiency of the spin injection/detection (with average contact spin polarization of ∼ 40%), confirms that the tunnel barrier is bilayer hBN[2]. The average resistance area product of the contacts on hBN/MoSe$_2$/Gr is about 4.5 kΩ·µm$^2$ at RT and about 64 kΩ·µm$^2$ at 75K.

## S3. Local magnetoresistance measurements

Four-terminal local magnetoresistance (MR) is measured as a function of out-of-plane magnetic field ($B_z$) to determine the behavior of the background signal. The measurement done on the pristine Gr channel (Figure S3a), shows a peak and a dip at low magnetic fields, representing the contribution of spin signal in the parallel and antiparallel configuration of the contacts. Therefore, in this local four-terminal, we are measuring both charge and spin transport due to the high spin polarization of the contacts. The local magnetoresistance measurement done on MoSe$_2$/Gr (Figure S3b), shows a different dependence of resistance versus magnetic field than what is shown in the main manuscript (Figure 2b). This behavior of magnetoresistance in the MoSe$_2$/Gr channel, clearly confirms that the observed change in the non-local resistance cannot be related to the background signals. The asymmetric component to the measured MR, can be attributed to the longitudinal Hall effects[12], most probably caused by the effect of the bubbles on the charge transport in the sample.

## S4. Models used for extraction of spin-transport parameters

### Uniform model

In this work, we have fitted the Hanle precession curves using the solution of the time-independent Bloch equations for a uniform system with spin lifetime anisotropy[7,13,14]:

$$D_s \frac{d^2 \mu_{sx}}{dx^2} - \frac{\mu_{sx}}{\tau_{\parallel}} + \gamma B_y \mu_{sz} - \gamma B_z \mu_{sy} = 0$$

$$D_s \frac{d^2 \mu_{sy}}{dx^2} - \frac{\mu_{sy}}{\tau_{\parallel}} + \gamma B_z \mu_{sx} - \gamma B_x \mu_{sz} = 0$$

$$D_s \frac{d^2 \mu_{sz}}{dx^2} - \frac{\mu_{sz}}{\tau_{\perp}} + \gamma B_x \mu_{sy} - \gamma B_y \mu_{sx} = 0$$

Where $\vec{\mu}_s = (\mu_{sx}, \mu_{sy}, \mu_{sz})$ is the 3 dimensional spin accumulation, $D_s$ is the spin diffusion coefficient, $\tau_{\parallel}$ and $\tau_{\perp}$ are the in- and out-of-plane spin relaxation times, respectively and $\gamma B = g\mu_B B/\hbar$ is the Larmor frequency with g the Landé factor, $\mu_B$ the Bohr magneton and $\hbar$ the reduced Planck constant. In our devices, the ferromagnetic contacts go all the way across the channel. This makes the spin accumulation, constant over the sample width ($W_s$) and allows us to base our analysis on the 1 dimensional case (x-dependent).

Now we add the effect of a Co spin injector (with a bilayer hBN tunnel barrier) that injects the spin current of $I_s = I \cdot P_{inj}$, polarized in the y direction (determined by its shape anisotropy), where I is the charge current and $P_{inj}$ is the polarization of the injector contact. This can be done by introducing a discontinuity of $IP_{inj}$ magnitude in the spin current $I_s = W_s/(eR_{sq}) d\mu_s/dx$ where $W_s$ is the width of the graphene, $R_{sq}$ is the square resistance of the graphene channel and e is the electron charge.

When the spin injector is placed at $x = 0$, if we apply a B field in the z direction, the spins precess in the $x - y$ plane and the spin accumulation $\mu_{sy}$ at $x = L$ is:

$$\mu_{sy} = \frac{P_{inj} eI R_{sq}}{2W_s} \text{Re} \left[ \frac{e^{-L\sqrt{(D_s \tau_\parallel)^{-1} - \frac{i\gamma B}{D_s}}}}{\sqrt{(D_s \tau_\parallel)^{-1} - \frac{i\gamma B}{D_s}}} \right]$$

This spin accumulation causes a nonlocal voltage that can be detected by a contact with spin polarization $P_{det}$, and can be normalized to obtain the nonlocal resistance $R_{prec} = \mu_{sy} P_{det}/eI$ [7].

The solution given above is with the assumption that the contact magnetizations are pointing in the y direction during the whole Hanle sweep. However, when the field is high enough, the magnetization of contacts can be pulled to the z direction by the magnetic field ($B_z$). In this case, the total measured spin signal will be:

$$R_{nl} = R_{prec} \cos(\theta_{inj}) \cos(\theta_{det}) + \sin(\theta_{inj}) \sin(\theta_{det}) \frac{P_{in} P_{det} R_{sq} \sqrt{D_s \tau_\perp}}{2W_s} e^{-L\sqrt{D_s \tau_\perp}} \quad [S3]$$

where $\theta_{inj}(\theta_{det})$ are the angle between the magnetization of the injector (detector) and the y direction.
In our fitting functions we have used equation S3 assuming that the contacts behave according to the Wohlfarth-Stoner model under a perpendicular magnetic field. This means that the z component of the contact magnetization is linear with the applied magnetic field ($M_z = MB/B_{sat}$, where M is the contact magnetization) until the saturation field $B_{sat}$ is reached and the contact magnetizations point to the B direction[15].
The model shown above has been used to fit the Hanle curves obtained in the pristine graphene regions with both magnetic fields in-plane($B_x$) and out-of-plane($B_z$). However, in our device there is a region covered by TMD. As a consequence, the spin relaxation time extracted from a fit to equation S3 will be affected by the TMD-covered region. To take that into account, we call the spin lifetimes extracted from this analysis 'effective' spin lifetimes.

**Four-region model**

To obtain the spin transport properties of the different regions in the sample from the 'effective' spin transport parameters, extracted by the fits to the uniform model, we have used two different models that account for the regions with different transport properties:

- Isotropic model

When the magnetic field is applied perpendicular to the graphene plane ($B_z$), as described above, the spin precession occurs in the $x - y$ plane and the spin lifetime anisotropy does not affect the spin precession.

In this case, the anisotropy becomes relevant only when $B_z$ is high enough to pull the contact magnetization out of plane. The signal in this field range is ruled by the spin lifetime in the z direction.

In this case, we have created a model that accounts for our device geometry (Figure S4a). In particular, our model has 4 different regions (I, II, III and IV) that are assembled in the following way: Region I is in the left side of the injector, placed at $x = 0$. Region II is at the right side of the injector and it connects the region under the spin injector with the TMD-covered region. Region III is the Gr channel covered with TMD and has different spin and charge transport parameters than the others. Finally, region IV is at the right side of the TMD-covered region and it has the same spin transport properties of the other outer regions. In conclusion: Regions I, II, and IV have same square resistance ($R_{Gr}$), diffusion coefficient ($D_{Gr}$), and spin lifetime ($\tau_{Gr}$). Region III has a square resistance ($R_{TMD}$), diffusion coefficient ($D_{TMD}$), spin lifetime out-of-plane ($\tau_{TMD}^{\perp}$) and spin lifetime in-plane ($\tau_{TMD}^{\parallel}$).

We match the adjacent regions, by using the following solutions of the Bloch equations:

$$\mu_{si} = A_i e^{x/\lambda\sqrt{1+i\gamma B \tau_r}} + B_i e^{x/\lambda\sqrt{1-i\gamma B \tau_r}} + C_i e^{-x/\lambda\sqrt{1+i\gamma B \tau_r}} + D_i e^{-x/\lambda\sqrt{1-i\gamma B \tau_r}}$$

Where i is the direction of the spin accumulation (i.e. x or y) and $\tau_r$ is the spin lifetime of region r. ($\tau_{Gr}$ for the pristine graphene and $\tau_{TMD}^{\parallel}$ for the TMD-covered part). To connect the different regions we use the following boundary conditions:

1. The spin accumulation $\mu_{si}$ is continuous in all the junctions.
2. The spin currents defined as $I_{si} = W_s/(eR_{sq})d\mu_{si}/dx$ are continuous everywhere apart from the spin injector, where there is a discontinuity of $IP_{inj}$.
3. The spin accumulation vanishes at $x \to \pm\infty$.

Using these 3 conditions[16,17] we obtain a system of 12 equations with 12 unknown parameters $A_i - D_i$ that we solve to obtain the spin signal at $x = L$.

- ### Anisotropic model

When we apply a magnetic field in the x direction and measure the spin signal across the TMD-covered region, the low-field spin precession data is strongly affected by the spin lifetime anisotropy. In this case, to solve the Bloch equations in a four-region model that accounts for the anisotropy of region III, we have used a finite difference calculation that implements an implicit Runge-Kutta method (Matlab). By using the same boundary conditions as shown in the previous section and including the effect of the contacts being pulled in the magnetic field direction, using the Wohlfarth-Stoner method, we can extract Hanle precession curves such as the one in Figure 4b of the main manuscript.

**S4.1. Hanle precession in pristine Gr with out-of-plane magnetic field ($B_z$)**

Here we compare Hanle spin precession in pristine Gr, measured by 3 detectors with different distances from the TMD/Gr region (Figure S4a). As the detector contact gets closer to the TMD/Gr region, we observe a considerable increase in the ratio of out-of-plane spin signal versus that of in-plane (Figure S4b). The parameters extracted by fitting the experimental data with the uniform model (Figure S4c and d), are reported in Table S1, showing an increase in spin diffusion coefficient and decrease in spin-relaxation time as the detector contact gets closer to TMD/Gr region (The estimated charge diffusion coefficient from charge transport is $D_c = 0.032 \, m^2/s$). This observation confirms the role of TMD/Gr region as a spin-sink only for the in-plane spins.

We can simulate the Hanle precession curves with the four-region Isotropic model and account for the spin sinking effect. We use the spin lifetimes, extracted by the uniform model for the detectors $V_{C6}$ (L(C1_C6) = 0.7 μm) and $V_{C3}$ (L(C1_C3) = 2.1 μm), as the preliminary parameters for generating the Hanle curves (Black line+symbol in Figure S4 e and f). This simulation is done by using the charge transport parameters, reported in Table S2. Then by fitting the obtained curves with equation S3 (uniform model), we estimate that if $\tau_{Gr} = 180$ ps and $\tau_{TMD}^{\parallel} = 40$ ps, the spin lifetime extracted for L = 0.7 μm and L = 2.1 μm is 171 ps and 154 ps, respectively (red lines in Figure S4 e and f). These values are in perfect agreement with the values obtained from the fitting of the experimental data. Thus we conclude that, 180 ps is a good estimate for the spin lifetime of the pristine Gr regions.

**S4.2. Hanle precession across the TMD/Gr region with out-of-plane magnetic field ($B_z$)**

We obtain an estimate for the spin lifetime anisotropy using the Hanle curves measured with $B_z$ across the TMD/Gr region. The behavior of $R_{nl}$ at small $B_z$ can be due to non-homogeneity of the Co contacts (since the out-of-plane spin signal is much larger than the in-plane signal, a small tilt of the contact magnetization in the z direction at zero field could have considerable effect on the measured data). This means that we cannot fit the Hanle curves of Figure 2b (main manuscript) to the Bloch equations to extract the in-plane spin lifetime. However, comparing the spin valve with the Hanle precession experiments at the highest B fields, when the contacts are close to the out-of-plane saturation, we see that the spin signal out of plane is 900 times larger than the in plane spin signal (see the main manuscript). We can use this value to estimate the ratio between the in- and out-of-plane spin lifetime assuming that the contact magnetizations are completely pulled in the z direction at the highest field of 1.2 T. In this case, we use the four-region model again. Note that in our measurement setup, $B_z$ is limited to 1.2 T, which is not sufficient for complete out-of-plane saturation of the contact magnetization. Therefore, the reported magnitude for the out-of-plane spin signal is the lower bound of the real value.

Since we do not have any accurate estimate of $\tau_{TMD}^{\perp}$, we take a range between 3 ps and 2 ns that safely covers the full window of the realistic values. For every value of $\tau_{TMD}^{\perp}$, we find the $\tau_{TMD}^{\parallel}$ that gives in-plane spin signal 900 times smaller than the out-of-plane spin signal (based on the result shown in Figure 2b, main manuscript). This analysis is done by applying the parameters shown in Table S2 for the geometry shown in Figure S5a, where the edge of the TMD-covered region is placed at 0.8 μm away from the spin injector and the distance between injector(C4) and detector(C5) is L = 4.1 μm. We see that the in-plane spin lifetime obtained using this method has to be shorter than $2.0 \pm 0.1$ ps, where the error comes from an uncertainty of 50 ps in the spin lifetime of the outer regions. The relatively high uncertainty in the spin lifetime of Gr (50 ps) is considered in order to confirm that our analysis is robust to the uncertainties of the spin transport parameters.

In Figure S5b and c, we can see that $\tau_{TMD}^{\parallel}$ has a significant dependence on $\tau_{TMD}^{\perp}$ when it is shorter than 250 ps. However, when $\tau_{TMD}^{\perp}$ increases beyond this value, $\tau_{TMD}^{\parallel}$ and the contact polarization show very small changes with respect to $\tau_{TMD}^{\perp}$. This is caused by the fact that the region covered with TMD is 2 μm long and the diffusion time across this part of the sample is of $\tau_d = \frac{L_{TMD}^2}{2D_{TMD}} = 38$ ps and this limits our resolution to extract $\tau_{TMD}^{\perp}$. However, the gray interval in Figure S5b which is related to contact polarization, determined by fittings to the Hanle measurements in the outer region, restricts the values of $\tau_{TMD}^{\parallel}$ and $\tau_{TMD}^{\perp}$ and gives a rough estimation of the spin lifetime anisotropy of $\tau_{TMD}^{\perp}/\tau_{TMD}^{\parallel} \sim 40 - 120$.

In order to do the Hanle precession measurements with $B_z$ and later on $B_x$, we had to move the sample to another measurement setup (setup1 with $B_z$ and setup2 with $B_x$) and we could not measure the Dirac curve across the TMD/Gr region in setup1. Comparison of the charge transport measurements in pristine graphene shows that the doping of the sample has changed. A fit of the conductivity of the pristine Gr channel versus gate voltage to $\sigma = \mu c(V_g - V_{CNP})$, allows us to estimate the position of the charge neutrality point ($V_{CNP}$). The comparison of the $V_{CNP}$ obtained in the both setups for Gr shows a shift of 25 V. Therefore, assuming that the mobility of TMD/Gr region has not changed, we obtained the value of $D_{TMD}= 0.053$ m$^2$/s and $R_{sq}= 352$ Ω used for the analysis mentioned above (Table S2, resulting in the curves of Figure S5). This estimation can be quite rough, therefore we test the robustness of the analysis considering a range of $D_{TMD}=$ 0.018 to 0.056 m$^2$/s and $R_{sq}=$ 313 to 2442 Ω, respectively, which covers the full interval of charge transport parameters measured in setup2. For this range of Dc for the TMD/Gr region the estimated value of the in-plane spin lifetime must be shorter than 3.5 ps. In conclusion, the analysis done for extraction of the range of the in-plane and out-of-plane spin lifetimes in TMD/Gr region in setup1 (with $B_z$), is only an estimation that is completed by the further measurements in setup2 (with $B_x$) that confirms the results of this analysis.

### S4.3. Hanle precession in pristine Gr region with in-plane magnetic field ($B_x$)

Here we show the Hanle precession curves obtained by applying $B_x$. The injector (C1) is placed at a distance of 3.8 µm from the edge of the TMD/Gr region and the detector C6 and C3 are placed at a distance of L(C1_C6)= 0.7 µm and L(C1_C3)= 2.1 µm from the injector (C1), respectively (Figure S6a). To obtain the spin transport parameters of the different regions we use the following procedure: We fit the Hanle curves to the Bloch equations for the uniform system and extract the effective spin lifetimes (Figure S6b and c). Here, we obtain $\tau_{eff} = 210 \pm 8$ ps for L(C1_C6) and $\tau_{eff} = 160 \pm 50$ ps for L(C1_C3), where the errors are determined from the fitting procedure. The large error bars, extracted from this analysis, do not allow us to determine the effect of the TMD/Gr region in a very accurate way. However, we can still use this model to confirm that the different measurements (with $B_z$ and $B_x$) create a consistent picture and our approximations are justified. To do so, we use the four-region model to simulate the Hanle curve in this geometry (using the charge transport parameters, reported in Table S3) and then we fit the modelling result with the uniform model and extract the effective spin lifetime. As a result, we obtain an effective lifetime of 191 ps for L=0.72 µm and 180 ps for L=2.06 µm (Figure S6d and e), indicating that $\tau_{TMD}^{\parallel} = 2$ ps and $\tau_{Gr} = 210$ ps are good estimates for our device. We found that this result is not very sensitive to $\tau_{TMD}^{\parallel}$ variations within the range from 2 to 5 ps. The other parameters are shown in Table S3.

### S4.4. Hanle precession across the TMD/Gr region with in-plane magnetic field ($B_x$)

Here we analyze our Hanle measurements with $B_x$ across the TMD/Gr with the four-region model (Figure S7a). To extract the correct $\tau_{TMD}^{\parallel}$, it is relevant to subtract the correct background resistance (-2.48 Ω). The saturation fields in x direction are between 0.15 T and 0.35 T. In this case, the spin signal at high B field, when compared with its value in the maxima/minima of the Hanles (at $B_x$= 0.1 T), informs us about the spin lifetime anisotropy in the system. When the distance between the injector (C5) and the detector (C2) is 5.6 µm, we observe the Hanle curves shown in Figure S7b. In this plot, we show the Hanle curves, calculated using the four-region model that includes the spin lifetime anisotropy of the TMD/Gr region with the parameters from Table S3. To obtain $\tau_{TMD}^{\perp}$, we have used the fact that the spin polarizations extracted from Hanle precession in this device are between 35 and 45%. This allows us to restrict the range of $\tau_{TMD}^{\perp}$ to 20 ps to 60 ps (Figure S7c). This implies that $\tau_{TMD}^{\parallel}$ has to be between 3 and 4 ps (Figure S7d) and therefore the spin lifetime anisotropy $\tau^{\perp}_{TMD}/\tau^{\parallel}_{TMD}$ has to be between 6.7 and 15. That gives us a value of $\tau^{\perp}_{TMD}/\tau^{\parallel}_{TMD}= 10.9 \pm 4$, considering $\tau^{\perp}_{TMD} = 40 \pm 15$ ps and $\tau^{\parallel}_{TMD} = 3.5 \pm 0.5$ ps.

We measure Hanle precession with a different contact configuration in which contact C4 is the injector and C5 is the detector (L(C4-C5)= 4.1 µm). We observe that the resulting Hanle curves (Figure S8a) have different shapes compared with the ones shown in the main manuscript (Figure 3b). This behavior can be explained by considering the contribution of the outer contacts. If we take into account the fact that the polarization of contact C4 is smaller than the polarization of one of the outer contacts, the contribution of the outer contact becomes relevant. We can understand some of these features, considering saturation of the inner contact (C4) at $B_x$ = 0.18 T. This saturation field is known from two-terminal in-plane Hanle precession measurements and indicates that the extra features at higher fields are caused by an outer contact which is not completely saturated in this range.

For the case of the injector C5 and detector C2 (Figure S8b), the relative contribution of the outer contact becomes smaller which indicates that the polarization of the contact C2 is higher than that of C4. However, the Hanle measurement still shows four levels that are slightly different, which is again attributed to the contribution of the mentioned outer contact. Here we only consider the intermediate levels (red and blue curves, also shown in Figure 3b of the manuscript), analysis of which with the four-region anisotropic model gives the lower bound for the lifetime anisotropy.

## S5. Gate dependence of spin signal in the monolayer MoSe$_2$/Gr heterostructure

Spin-transport in the pristine Gr channel is measured at different gate voltages ($V_g$) with doped Si as back-gate, using 300 nm SiO$_2$ (not shown here). The magnitude of the spin-signal in Gr follows the variation of the channel resistance as a function of $V_g$. By tuning the $V_g$ from 0 V to -60 V (towards the neutrality point), we observe about 2 times increase in the spin signal as the Gr channel resistance increases by 3 times.

Moreover, we evaluate the modification of spin-transport across the TMD/Gr region when the gate voltage is changed. The band alignment between Gr and MoSe$_2$ is expected to bring the Fermi-level in the bandgap of MoSe$_2$. By tuning the electric field applied by the gate voltage, we can potentially shift the position of Fermi level in the band structure of MoSe$_2$. However, experimentally, in the range of $V_g$ from -40 V to +40 V, we do not observe considerable change in the magnitude of the in-plane spin signal which rules out the TMD spin absorption as the spin relaxation mechanism in our sample. Namely, in Figure S9, we show the anisotropic Hanle precession measurements with $B_x$, measured across TMD/Gr region at different gate voltages. We observe that the ratio of the spin signal at $B_x$ = 0 T with respect to the spin signal at $B_x$ = 0.1 T (generated by precession of spins in the y-z plane), changes less than a factor of 2 in the full range of gate voltages. These measurements together with the fact that the spin resistance in monolayer TMD is much larger than that of Gr, confirms that the spin transport in the monolayer TMD/Gr heterostructure is modulated by the proximity effect of the TMD but not by spin absorption.

## S6. Anisotropic spin transport measurement on monolayer WSe$_2$/Gr heterostructure

We observe similar behavior of the nonlocal magnetoresistance versus $B_z$ in a monolayer WSe$_2$/Gr heterostructure in similar device geometries. Here we show the optical microscope and AFM images of the Gr flake, fully covered by monolayer WSe$_2$, on a SiO$_2$(300 nm)/Si substrate (Figure S10a). The results of spin-transport at 75 K in the fabricated device with TiO$_2$/Co contacts are shown in Figure S10b. These measurements demonstrate a very small magnitude of in-plane spin signal of about 0.9 mΩ, detected over a 1.5 µm WSe$_2$/Gr channel in the conventional spin-valve geometry. This spin-valve measurement shows only the switch of magnetization of only one of the inner contacts (injector or detector), meaning that the magnetization of the other contact is pinned and does not change by the magnetic field along the Co easy axis.

Hanle measurements show that as $B_z$ increases and the magnetization of the Co contacts starts getting pulled out of Gr plane, the $R_{nl}$ signal increases due to the contribution of out-of-plane spins. This observation is in agreement with our results on monolayer $MSe_2$/monolayer Gr reported in the main text. Since the graphene channel is fully covered with $WSe_2$, we can apply the uniform model to extract the spin-transport parameters. The fit to the solution of Bloch equations for the Hanle precession (subtraction of parallel and antiparallel curves ($\Delta R_{nl}/2$), shown in the inset of Figure S10b) gives a value of about 11 ps for the in-plane spin lifetime. By extrapolation of the $R_{nl}$ signal to the saturation magnetic field of $B_z$ = 1 T, we estimate the spin lifetime anisotropy of $\tau_\perp/\tau_\parallel \approx 40$. Considering the relation[18] of $\tau_\perp/\tau_\parallel \approx (\lambda_{VZ}/\lambda_R)^2$ and the values of $\lambda_{VZ} = 1.2$ meV and $\lambda_R = 0.56$ meV for $WSe_2$/Gr[19], we obtain a relatively strong intervalley scattering of $\tau_{iv} \sim 8.7\,\tau_p \approx 0.47$ ps (with $\tau_p = 0.054$ ps) in this system.

# List of Figures

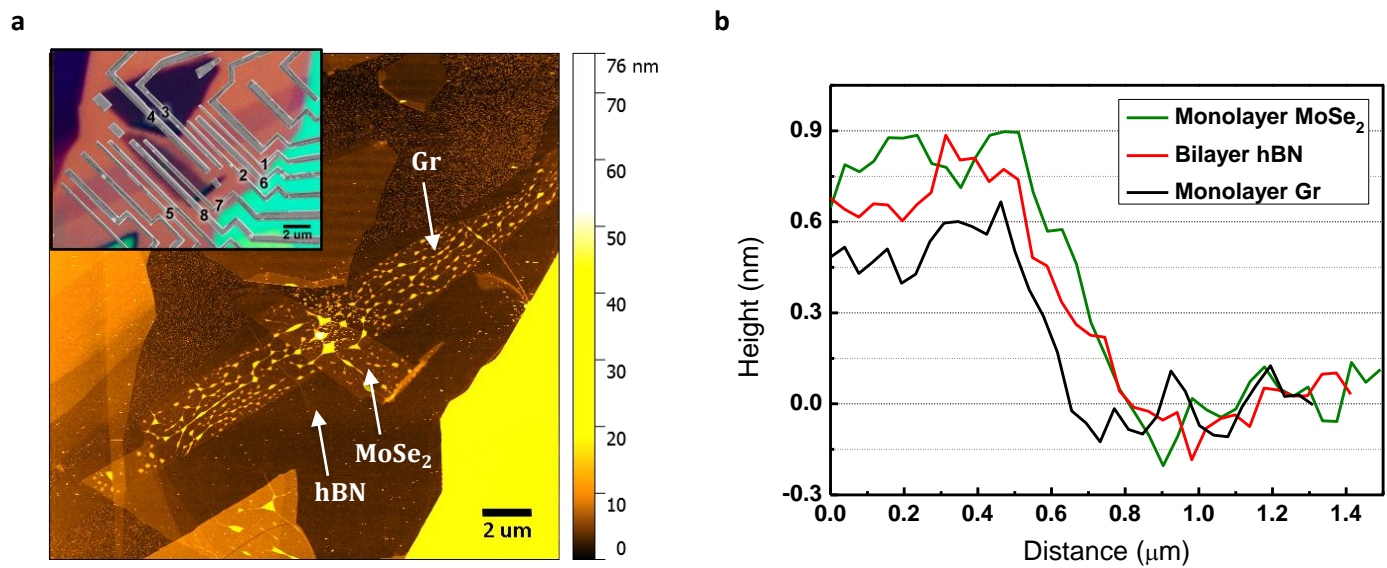

**Figure S1| Atomic force microscope (AFM) characterization of the vdW heterostructure of bilayer hBN/monolayer MoSe$_2$/monolayer Gr on SiO$_2$/Si substrate.** (a) AFM image of the fabricated stack. The combination of optical microscope image of the vdW stack and an SEM image of electrodes is shown in the inset. The electrodes used for the measurements are numbered. (b) Comparison of the height profiles of each layer.

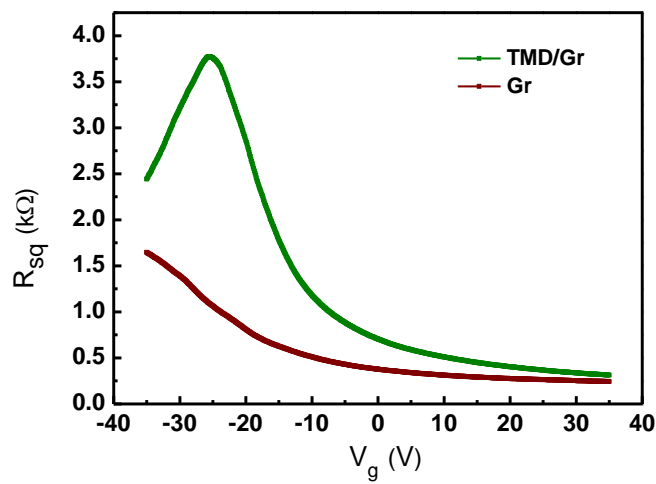

**Figure S2| Charge transport measurements in Gr and TMD/Gr regions.** Square resistance ($R_{sq}$) of the Gr channel as a function of gate voltage, measured by local four-terminal geometry on pristine Gr and across MoSe$_2$-covered Gr at 75 K. To obtain the $R_{sq}$ of the TMD/Gr region, with use the voltage probes of C4 and C5 and we subtract the contribution of the bare Gr regions between the contacts. Contacts are numbered in Figure S1a.

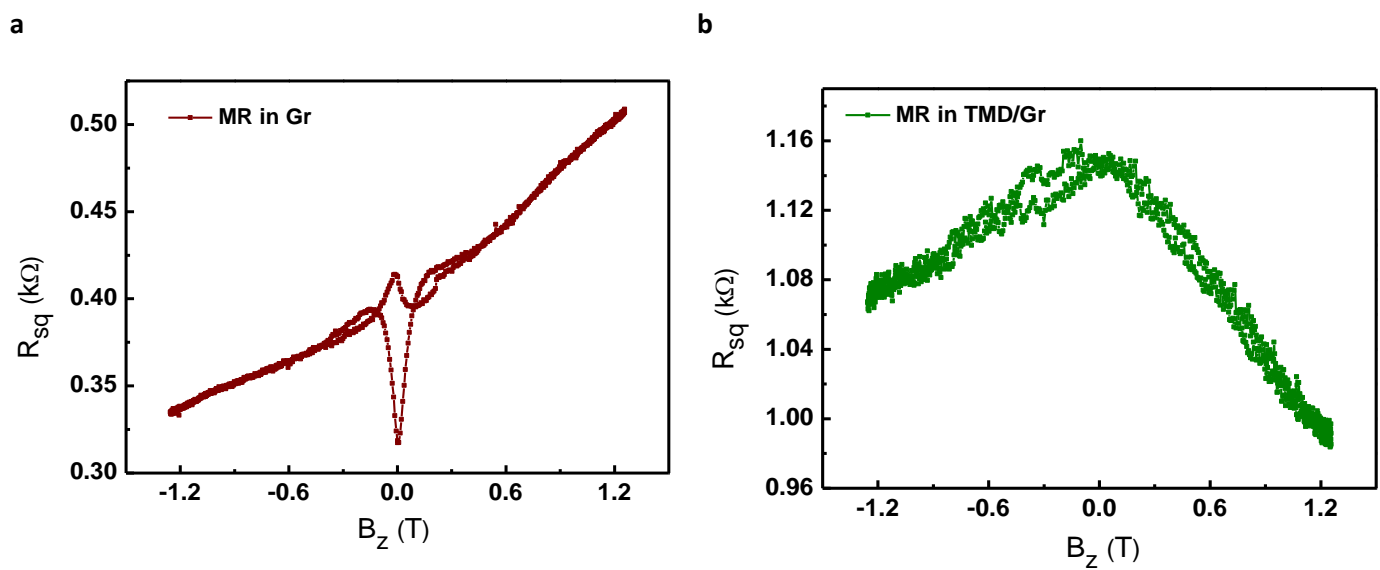

**Figure S3| Local magnetoresistance measurements at 75 K and $V_g = 0$ V.** $R_{sq}$ measurement as a function of $B_z$, measured (a) on the pristine Gr channel (with C1 and C6 as the voltage probes) and (b) on the MoSe$_2$-covered Gr region (with both voltage probes on TMD, C7 and C8). Contacts are numbered in Figure S1a.

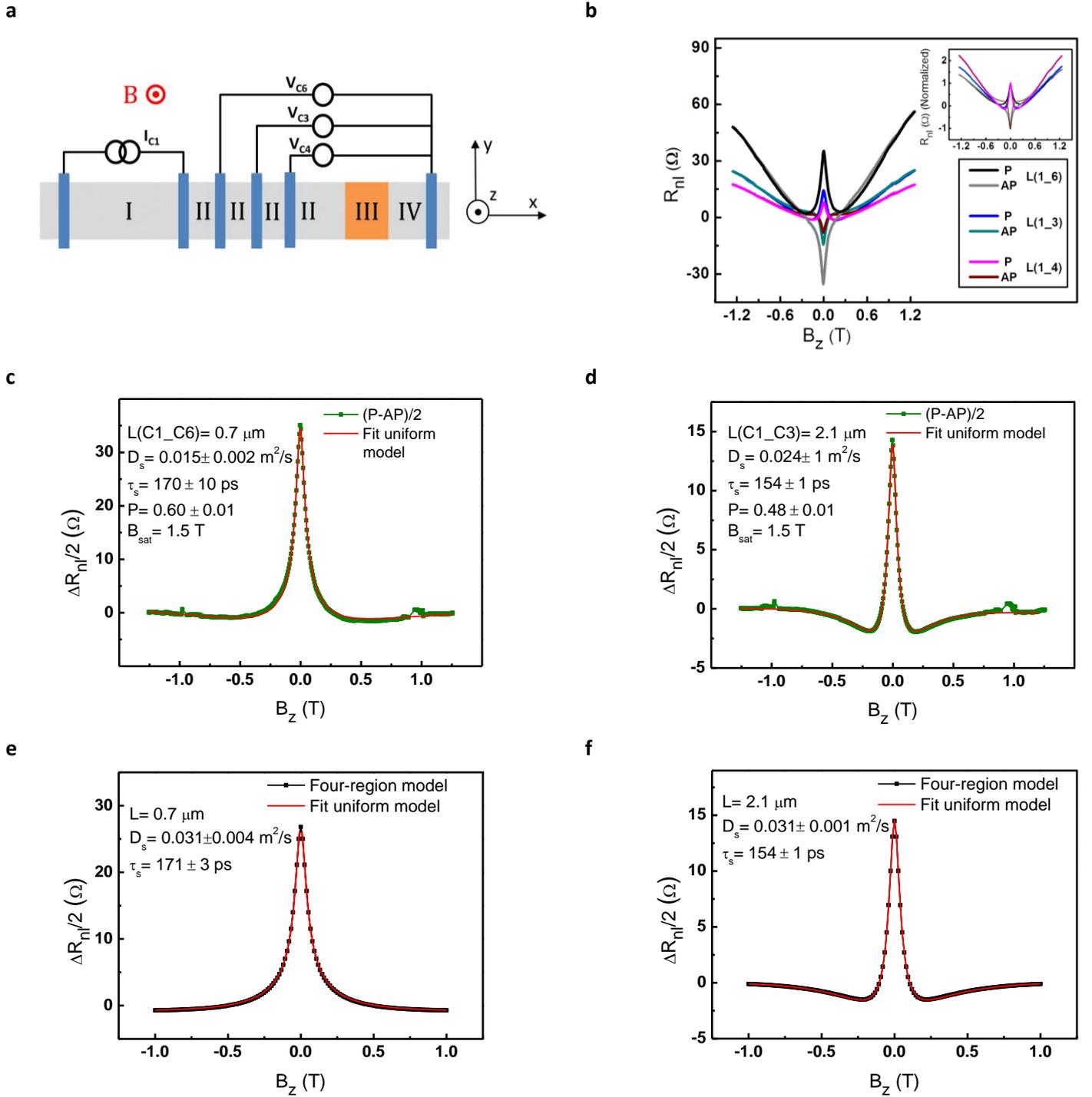

**Figure S4| Hanle precession in pristine Gr with $B_z$.** (a) Schematic picture of the modelled device geometry (Contacts are numbered according to Figure S1a). (b) $R_{nl}$ measurement, done with contact C1 as the spin injector and contact C6, C3 and C4 as the detectors, with injector-detector distances of L(C1_C6)= 0.7 μm, L(C1_C3)= 2.1 μm and L(C1_C4)= 2.9 μm. The inset shows the normalized value of $R_{nl}$ with respect to the detected signal at B= 0 T. This comparison is showing an increase in ratio of out-of-plane/in-plane spin-signal as the detector contact gets closer to TMD/Gr region. (c) and (d) are $\Delta R_{nl}/2 = (R_p - R_{ap})/2$ as a function of $B_z$ with the fit to the uniform model for measurement done on L(C1_C6) and L(C1_C3), respectively. (e) and (f) are the precession curves simulated by the four-region model and using the parameters shown in Table S2 and the fits with the uniform model. The large polarization obtained in (c) is most likely caused by the underestimation of $D_s$ due to the fact that L(C1_C6) is shorter than the spin relaxation length. The errors of data points (small jumps in $R_{nl}$) at $\sim \pm 1$ T in (c) and (d) are related to a small displacement of the magnetic poles.

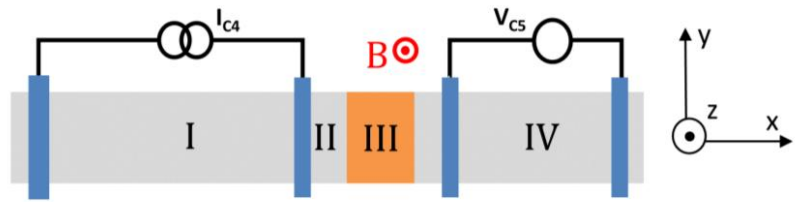

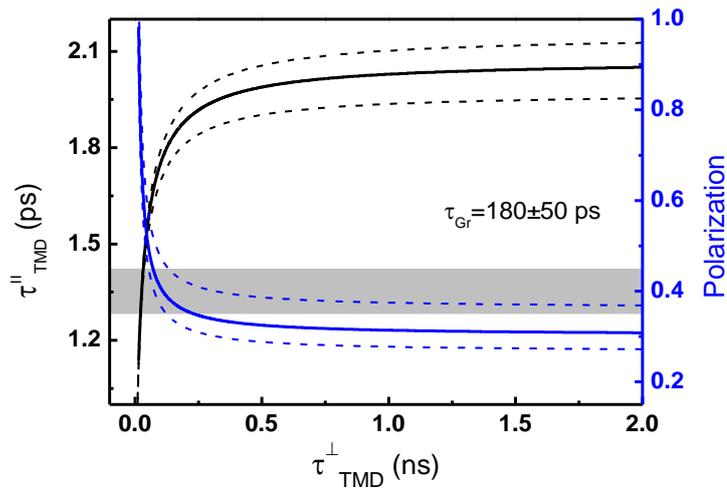
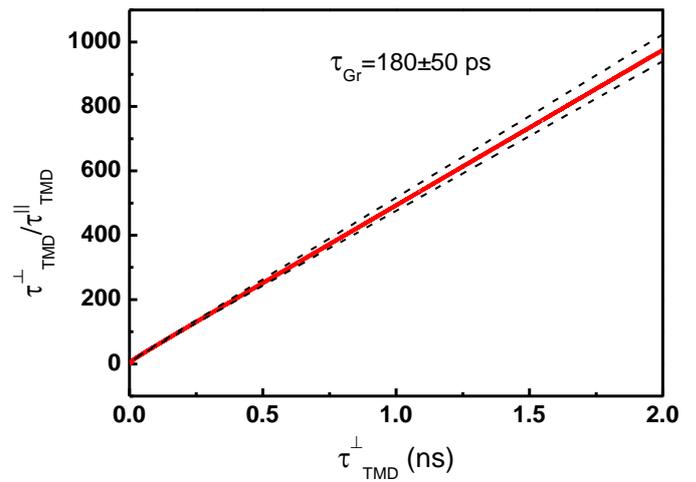

**Figure S5| Spin lifetime anisotropy from measurements across TMD/Gr with $B_z$.** (a) schematic picture of device geometry (Contacts are numbered according to Figure S1a). (b) In-plane spin lifetime ($\tau_{TMD}^{\parallel}$) and spin polarization as a function of the out-of-plane spin lifetime ($\tau_{TMD}^{\perp}$), estimated from in-plane and out-of-plane spin signal, measured with injector C4 and detector C5 (main manuscript Figure 2b). The spin polarizations of the Co contacts are about 35% to 45%, extracted from Hanles measured in the outer regions (shown as the gray interval). (c) Spin lifetime anisotropy in the system as a function of the out-of-plane spin lifetime. The dashed lines in (b) and (c) indicate the confidence interval associated with an uncertainty of 50 ps in the determination of $\tau_{Gr}$.

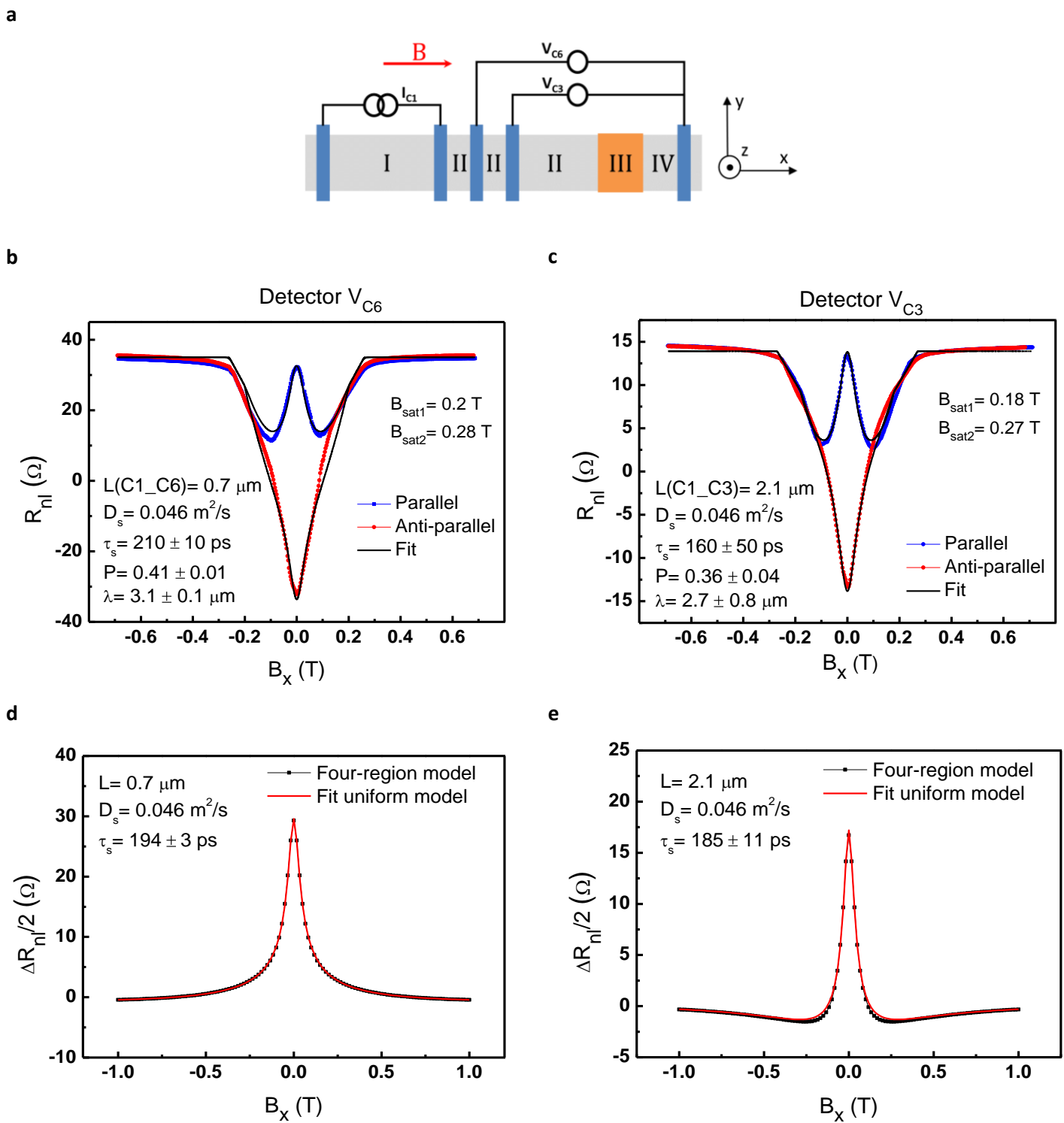

**Figure S6| Hanle precession with $B_x$ in Gr.** (a) Schematic illustration of the measurement geometry (Contacts are numbered according to Figure S1a). $R_{nl}$ measured as a function of $B_x$ and fits with four-region model, with C1 as injector and (b) C6 as the detector, and (c) C3 as the detector. The magnetic field for the saturation of contacts in x direction ($B_{sat}$) is considered about 0.2 T and 0.28 T for the fittings. (d) and (e) are the precession curves simulated applying the four-region model and using the parameters shown in the Table S3 and the fits with the uniform model.

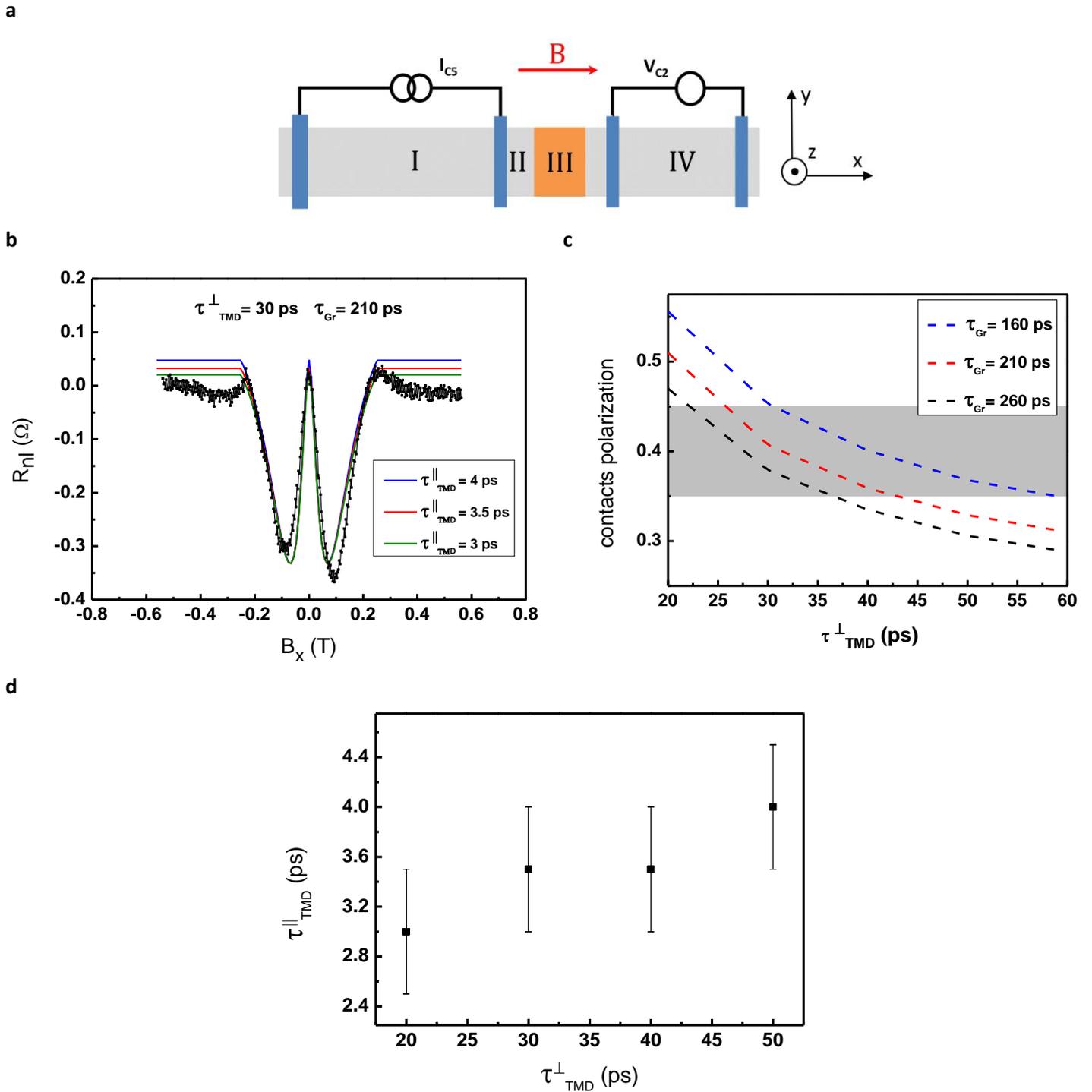

**Figure S7| Modelling of the Hanle precession across TMD/Gr with $B_x$.** (a) Schematic illustration of the measurement geometry (Contacts are numbered according to Figure S1a). (b) Data measured across TMD (with injector C5 and detector C2, and channel length of L(C5_C2)= 5.6 μm, covered with 2 μm TMD) and fit with four-region anisotropic model, for different $\tau^{\parallel}_{TMDC}$. (c) Contact polarization as a function of $\tau^{\perp}_{TMDC}$. The dashed lines represent the values obtained assuming an uncertainty of 50 ps in $\tau_{Gr}$. The gray interval shows the range of the contact polarization that can be considered for this device, deduced from Hanle measurements on pristine Gr region. (d) $\tau^{\parallel}_{TMDC}$ as a function of $\tau^{\perp}_{TMDC}$, the error bars correspond to the uncertainties in fitting the data to the model.

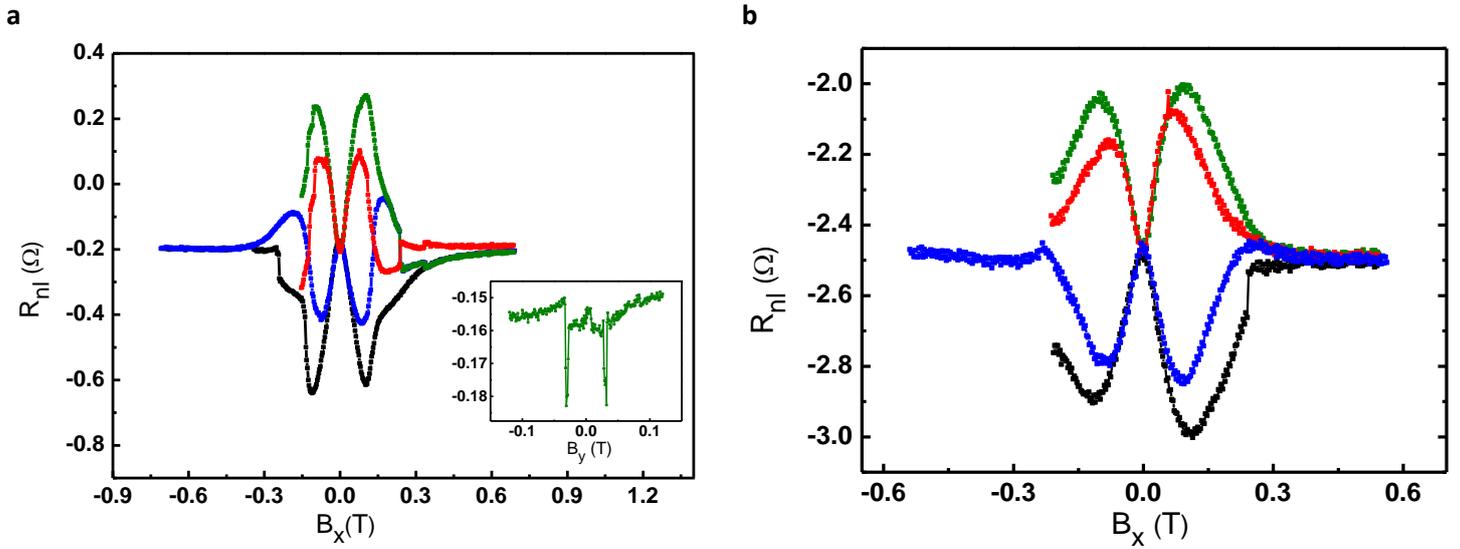

**Figure S8| Hanle precession across TMD/Gr region with $B_x$.** (a) $R_{nl}$ is measured as a function of $B_x$, with C4 as injector and C5 as detector. The contribution of the outer contact creates 4 levels of Hanle curves. The corresponding lateral spin-valve measurement is shown in the inset. (b) $R_{nl}$ is measured as a function of $B_x$, with C5 as the injector and C2 as the detector. Contacts are numbered in Figure S1a. In these plots background resistance is not subtracted.

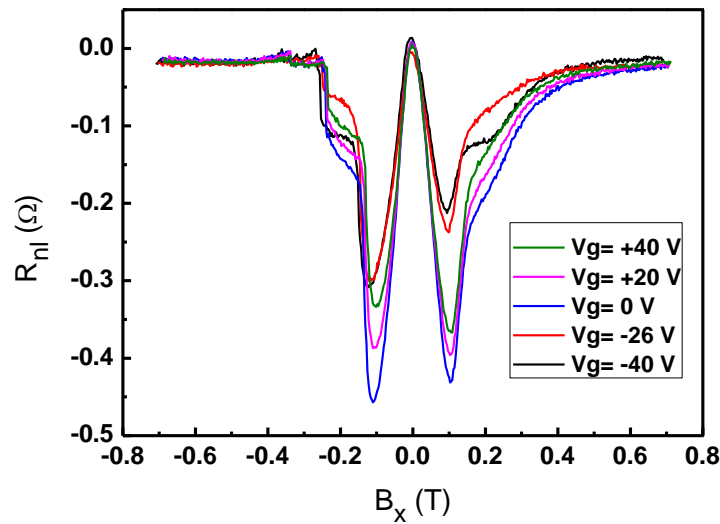

**Figure S9| Gate dependence of $B_x$ induced Hanle precession measurement across TMD/Gr.** Measurement of $R_{nl}$ as a function of $B_x$ at different back-gate voltages at 75 K. The distance between the spin injector (C3) and the spin detector (C5) is L(C3_C5)= 5 µm. Contacts are numbered in Figure S1a.

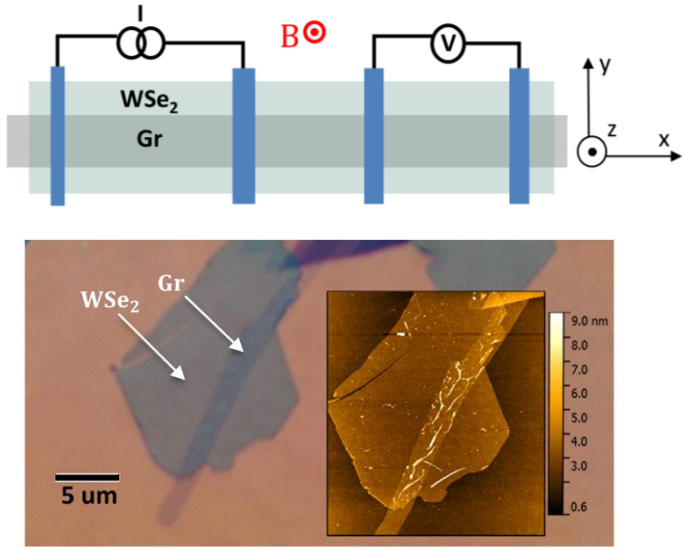
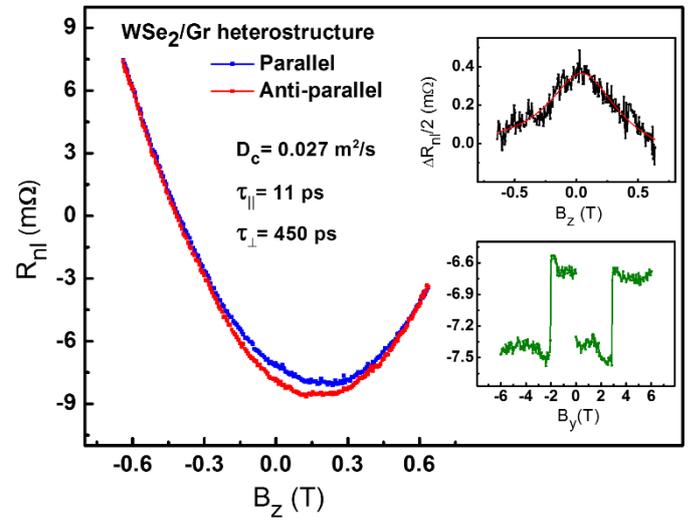

**Figure S10| Anisotropic spin transport measurement on a monolayer WSe$_2$/Gr heterostructure.** (a) Optical microscope and AFM image of a monolayer WSe$_2$/ monolayer Gr heterostructure and sketch of the device geometry, (b) Nonlocal magnetoresistance (R$_{nl}$) measurement as a function of B$_z$ in the device based on monolayer WSe$_2$/Gr on SiO$_2$(300 nm)/Si substrate with TiO$_2$/Co contacts. The corresponding nonlocal spin-valve measurement and the subtraction of parallel and antiparallel states of Hanle precession with the fit to uniform model are shown in the insets. The measurements are done with I$_{ac}$ = 100 μA and at Vg= +40 V for reduction of the noise level. The asymmetric component in the measured Hanle curves is attributed to longitudinal Hall currents caused by the nonhomogeneity of the TiO$_x$ barriers.

## List of Tables

**Table S1| Spin transport parameters, extracted from fits to the uniform model of the Hanle precession measurements, obtained in pristine Gr with $B_z$.**

| Detector No. | Distance of detector contact to TMD (µm) | $D_s$ (m²/s) | $\tau_s$ (ps) | out-of-plane/In-plane spin signal |
|---|---|---|---|---|
| 1 | 3 | 0.015 | 170 | 1.47 |
| 2 | 1.7 | 0.024 | 154 | 1.75 |
| 3 | 0.8 | 0.034 | 146 | 2.17 |

**Table S2| Parameters used to simulate the Hanle measurements in pristine Gr using the four-region model.** $R_{Gr}$, $D_{Gr}$ and $\tau_{Gr}$ respectively are the square resistance, charge diffusion coefficient and spin lifetime of regions I, II and IV. $R_{TMD}$, $D_{TMD}$ and $\tau_{TMD}^{\parallel}$ respectively are the square resistance, charge diffusion coefficient and in-plane spin lifetime of the TMD/Gr region with the length of $L_{TMD}$. The $L(inj - enc)$ is the distance from the spin injector to the TMD covered region.

| | |
|---|---|
| $R_{Gr}$ | 451 Ω |
| $R_{TMD}$ | 354 Ω |
| $D_{Gr}$ | 0.032 m²/s |
| $D_{TMD}$ | 0.053 m²/s |
| $\tau_{Gr}$ | 180 ps |
| $\tau_{TMD}^{\parallel}$ | 40 ps |
| $L_{TMD}$ | 2 µm |
| $L(inj\_enc)$ | 3.8 µm |

**Table S3| Parameters used to simulate the Hanle curves with $B_x$ using the four-region model in the geometry of Figure S6a.** $R_{Gr}$, $D_{Gr}$ and $\tau_{Gr}$ respectively are the square resistance, charge diffusion coefficient and spin lifetime of regions I, II and IV. $R_{TMD}$, $D_{TMD}$, $\tau_{TMD}^{\perp}$ and $\tau_{TMD}^{\parallel}$ respectively are the square resistance, charge diffusion coefficient out-of-plane and in-plane spin lifetime of TMD/Gr region (III). $L_{TMD}$ is the length of the TMD covered region and $L(inj - enc)$ is the distance from the spin injector to the edge of the TMD-covered region.

| | |
|---|---|
| $R_{Gr}$ | 377 Ω |
| $R_{TMD}$ | 706 Ω |
| $D_{Gr}$ | 0.046 m²/s |
| $D_{TMD}$ | 0.038 m²/s |
| $\tau_{Gr}$ | 210 ps |
| $\tau_{TMD}^{\perp}$ | 210 ps |
| $\tau_{TMD}^{\parallel}$ | 2 ps |
| $L_{TMD}$ | 2 µm |
| $L(inj - enc)$ | 3.8 µm |